# Structure and magnetic properties of melilite-type compounds $RE_2Be_2GeO_7$ (RE = Pr, Nd, Gd-Yb) with Rare-Earth ions on Shastry-Sutherland lattice

Malik Ashtar,[1,=] Yuming Bai,[1,†], Longmeng Xu[1], Zongtang Wan[1], Zijun Wei[1], Yong Liu[2], Mohsin Ali Marwat[3], Zhaoming Tian[1*]

[1] Wuhan National High Magnetic Field centre and School of Physics, Huazhong University of Science and Technology, Wuhan 430074, Peoples R China.

[2] School of Physics, Wuhan University, Wuhan 430072, Peoples R China.

[3] College of Materials Science and Engineering, Huazhong University of Science and Technology, Wuhan 430074, PR China

## ABSTRACT:

Rare-earth (RE) based frustrated magnets as typical systems of combining strong spin-orbit coupling, geometric frustration and anisotropic exchange interactions, can give rise to diverse exotic magnetic ground states such as quantum spin liquid (QSL). The discovery of new RE-based frustrated materials is crucial for exploring the exotic magnetic phases. Herein, we report the synthesis, structure and magnetic properties of a family of melilite-type $RE_2Be_2GeO_7$ (RE = Pr, Nd, Gd-Yb) compounds crystallized in a tetragonal $P\bar{4}2_1m$ structure, where magnetic $RE^{3+}$ ions lay out on Shastry-Sutherland lattice (SSL) within *ab*-plane and are well separated by nonmagnetic $[GeBe_2O_7]^{-6}$ polyhedrons along *c*-axis. Temperature (*T*)-dependent susceptibilities $\chi(T)$ and isothermal magnetization $M(H)$ measurements reveal that most $RE_2Be_2GeO_7$ compounds except RE=Tb show no magnetic ordering down to 2 K despite the dominant antiferromagnetic (AFM) interactions, where $Tb_2Be_2GeO_7$ undergoes AFM transition with Néel temperature $T_N$~ 2.5 K and field-induced spin flop behaviors (*T*< $T_N$). In addition, the calculated magnetic entropy change $\triangle S_m$ from the isothermal $M(H)$ curves reveal a viable magnetocaloric effect (MCE) for $RE_2Be_2GeO_7$ (RE =Gd, Dy) in liquid helium temperature regimes, $Gd_2Be_2GeO_7$ shows maximum $\triangle S_m$ up to 54.8 J $K^{-1}$ $Kg^{-1}$ at $\triangle H$= 7 T and $Dy_2Be_2GeO_7$ has largest value $\triangle S_m$=16.1 J $K^{-1}$ $kg^{-1}$ at $\triangle H$= 2 T in this family. More excitingly, rich diversity of RE ions in this family enables an archetype for exploring exotic quantum magnetic phenomena with large variability of spin located on SSL lattice.

# ■ INTRODUCTION

Geometrically frustrated magnets as a fertile playground to explore the exotic quantum magnetic phases, have attracted enormous interest in condensed matter physics. In such systems, magnetic moments are located on the geometric lattices usually constructed by connections of corner or edge-shared triangular or tetrahedra lattices,[1-6] where geometric lattice network can be in two-dimension (2D) like triangular, kagome and honeycomb type, or in three-dimension like pyrochlore, garnet, and hyperkagome type. For 2D frustrated systems with spin S=1/2, competing confined spin exchange interactions restrict the magnetic degrees of freedom, lead to strong frustration accompanied by enhanced quantum fluctuations. Ultimately, the system is prevented from adopting conventional long-range magnetic ordering but in favour of non-trivial magnetic states like quantum spin liquid (QSL) state,[2,7] typical examples include the 2D triangular lattice κ-(BEDT-TTF)$_2$Cu$_2$(CN)$_3$ [3,8] and 2D kagome lattice ZnCu$_3$(OH)$_6$Cl$_2$ and its derives, [9-11] which don't show long-range magnetic order or spin freezing with temperature down to milliKelvin temperatures as best candidate materials realizing QSL state. On the other hand, to unveil novel exotic magnetic states, extensive theoretical and experimental works have been dedicated to search for new frustrated materials on various lattices. A particular interesting structure in this respect is called Shastry-Sutherland lattice (SSL),[12] which consists of planes of orthogonal dimers with intradimer exchange $J_1$ and interdimer exchange $J_2$ as shown in Figure 1a. Since firstly proposed by Shastry and Sutherland as theoretical toy model,[12] it has received substantial attention to investigate novel quantum magnetic phenomena. In this SSL model, different magnetic ground states are predicted depending on the relative strength of $J_1$ and $J_2$. When $J_2/J_1$ is large, it will support the Heisenberg antiferromagnetic (AFM) order, small $J_2/J_1$ will induce a disordered dimer singlet ground state, and insulating SrCu$_2$(BO$_3$)$_2$ exemplifies this disordered "spin liquid (SL)" regime.[13-15] Moreover, intriguing quantum phase transition from AFM to SL phases can take place by tuning $J_2/J_1$ under the external magnetic field or high pressure.[16,17] New material realizations on SSL motif will offer promising platform to study the predicted magnetic phases.

Quite recently, frustrated magnets incorporating 4$f$ rare-earth (RE) ions attract intensive attention to explore exotic magnetic phases arising from the multiple interplay of spin-orbit entangled $j_{eff}$=1/2 moment, geometric lattice frustration, exchange dipolar interaction and diverse spin-types of RE$^{3+}$.[18-22] And indeed, a wide variety of exotic magnetic phases have been experimental identified, such as QSL,[7,22] quantum spin ice [18,23,24] and multipolar order,[25,26] etc. In terms of the SSL systems, metallic RE$_2$Pt$_2$Pb (RE= Yb, Ce)[27,28] and REB$_4$ (RE =Ce-Yb) [29,30] compounds have been well studied, some of interesting magnetic phenomena have been reported, as examples, field-induced quantum phase transition from AFM to SL phase observed in RE$_2$Pt$_2$Pb,[27] unusual magnetization plateaus phenomena under field found in REB$_4$ series.[30] On other side, insulating RE-based oxides are also interesting to unveil the intrinsic SSL physics fully dictated by topological lattice frustrations of RE$^{3+}$ ions, avoiding the influence of exchange interactions mediated by itinerant conduction electrons in metallic systems. But, the related materials are still rare, only $RE_2BaZnO_5$ (RE = Pr-Ho) and its derivatives are magnetically characterized,[31-33] restrict further explorations of novel magnetism on SSL occupied by $RE^{3+}$ ions. Therefore, discovering new RE-based materials with SSL lattice is highly desirable to uncover the exotic magnetic phenomena, motivating present study.

From the application standpoint, low dimensional RE-based frustrated magnets are attractive as magnetic refrigeration materials in liquid helium temperature regime, enhanced magnetocaloric (MC) effect can be obtained during the adiabatic demagnetization process, because the large magnetic entropy ($\Delta S_m$) arising from highly degenerate magnetic ground states at zero field.[34] Additionally, the lower temperature limit for magnetic cooling is determined by magnetic ordering temperature of magnetic reagents. In RE-based frustrated magnets, due to the multiple competing interactions, the suppressed or absence of magnetic ordering will be resultantly endowing them to achieve ultra-low temperatures. Moreover, the variety of spin types, large moments and magnetic anisotropy of $RE^{3+}$ ions make them flexible for designing MC materials working at different field regions. As typical examples, Gd-based frustrated magnets like commercial gadolinium gallium garnet $Gd_3Ga_5O_{12}$ (GGG) as Heisenberg-type antiferromagnets, can bear large MC effect at high fields ($\mu_0H > 5T$).[35] As comparisons, Ising-anisotropic $Dy_3Ga_5O_{12}$ exhibits better MC performance at low field regions ($\mu_0H < 2T$).[36]

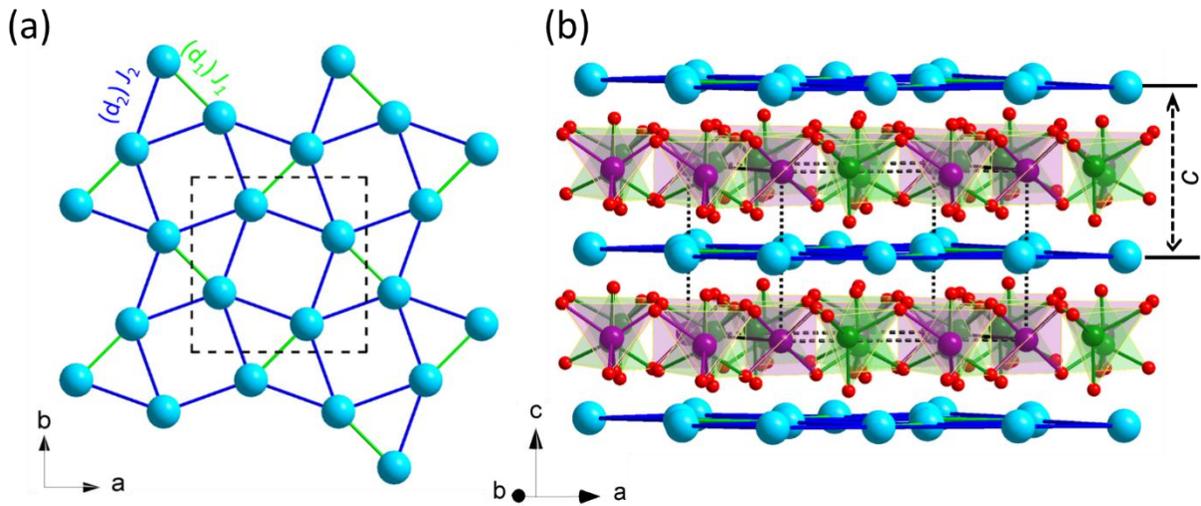

Figure 1. (a) The schematic view of Shastry-Sutherland lattice (SSL), the green and blue lines represent triangle and square bonds. (b) The stacking fashion of SSL layers incorporating of $RE^{3+}$ moments along $c$ axis in $RE_2Be_2GeO_7$, the $RE^{3+}$ layers are well separated by nonmagnetic $[GeBe_2O_7]^{-6}$ polyhedrons.

In this work, we report the magnetic properties of $RE_2Be_2GeO_7$ (RE = Pr, Nd, Gd-Yb) melilites comprising the frustrated SSL of $RE^{3+}$ moments. All $RE_2Be_2GeO_7$ compounds crystallize in tetragonal structure with $P\bar{4}2_1m$ space group, which were previously synthesized and structurally characterized firstly by Ochi *et al* and later by Mill *et al* from a structural point of view,[37,38] but topological lattice geometry on spin frustration has not been studied until now. Our structural analysis reveals that the $RE^{3+}$ ions form the SSL in the *ab* plane and magnetic SSL layers are stacked in an AAA-type fashion well separated by nonmagnetic $[GeBe_2O_7]^{-4}$ sheets along *c*-axis. Magnetic characterizations show most $RE_2Be_2GeO_7$ (RE = Pr, Nd, Gd-Yb) compounds except of RE = Tb show no magnetic order down to 2 K despite their dominant antiferromagnetic interactions, where Tb-analogue exhibits AFM transition with Néel temperature $T_N$~ 2.5 K. The calculated magnetic entropy change $\Delta S_m$ reveal that $Gd_2Be_2GeO_7$ has the largest MCE at high fields (5T $\leq H\leq$ 7 T) and $Dy_2Be_2GeO_7$ exhibits better MCE performance at low fields ($H \leq$ 2 T), as suitable materials for magnetic refrigeration in liquid helium temperatures.

■ **EXPERIMENTAL SECTION**

The series of $RE_2Be_2GeO_7$ (RE = Pr, Nd, Gd-Yb) polycrystals were synthesized by a solid state reaction method using $GeO_2$(99.99%), BeO(99.5%) and RE oxides $RE_2O_3$ (RE=Pr, Nd, Gd-Yb; 99.99%) as starting materials. Before using, raw materials $RE_2O_3$ (RE=Pr, Nd) were pre-dried for 12 hours at 800°C in argon atmosphere. Stoichiometric amounts of the starting materials with mole ratio of RE:Be:Ge=2:2:1 were weighted, ground in glove box. After that, the mixtures were pre-reacted in air at temperature 1100°C for 24 hours. For $RE_2BeGeO_7$(RE=Pr,Nd) samples, the products were re-ground and reacted at temperature of 1200°C-1250°C for 4 days with intermediate grindings. For RE=Gd-Yb compounds, higher reaction temperature of 1300°C is required to obtain pure phase sample.

The phase purity and crystal structure of $RE_2Be_2GeO_7$ were checked by room temperature powder X-ray diffraction (XRD, Rigaku Smartlab) with Cu Kα radiation (λ= 1.5418 Å). The Rietveld refinements of XRD data were performed for structural analysis using Material Studio software.[39] The magnetic properties were measured using a commercial SQUID magnetometer (MPMS, Quantum Design) and commercial Physical Property Measurement System (PPMS, Quantum Design) equipped with a vibrating sample magnetometer (VSM) option.

■ **RESULTS AND DISCUSSION**

**Structure descriptions.**

All serial $RE_2Be_2GeO_7$ (RE = Pr-Yb) melilites are isostructural and crystallize into tetragonal structure in space group $P\bar{4}2_1m$ (No. 113). In the unit cell, there are six crystallographic sites including one unique RE (Wyckoff site 4e), one Ge (Wyckoff site 2a), one Be (Wyckoff site 4e) and three types of O (Wyckoff site 2c,2a,8f) sites. Using the crystal structure of $Y_2Be_2SiO_7$ as a starting model,[40] the experimental and refined XRD patterns for selected $RE_2Be_2GeO_7$ (RE = Pr, Yb) compounds are shown in Figure 2a. The fitted profiles match the experimental data well giving reliability factors $R_p$ (2.95-3.91) and $R_{wp}$ (4.01-5.22). The obtained structural lattice parameters and atomic coordinates are summarized in Table S1 (see Supporting Information), the structural parameters are in accordance with previous report.[37,38] From the refinement, no significant anti-site mixing between magnetic $RE^{3+}$ and nonmagnetic $Be^{2+}/Ge^{4+}$ ions are detected, indicative of ordered arrangement of magnetic $RE^{3+}$ and nonmagnetic $Be^{2+}/Ge^{4+}$ cations. The free of site mixing can be related to the difference of ionic radii and coordination number between $RE^{3+}$ and $Be^{2+}/Ge^{4+}$ cations, and this is critical to investigate the intrinsic magnetic behaviors from the ordered $RE^{3+}$ moments, avoid the influence of structural disorder on magnetic ground state as the situation in $YbMgGaO_4$ where the site-mixing occupation of Mg and Ga atoms exist. [20,41]

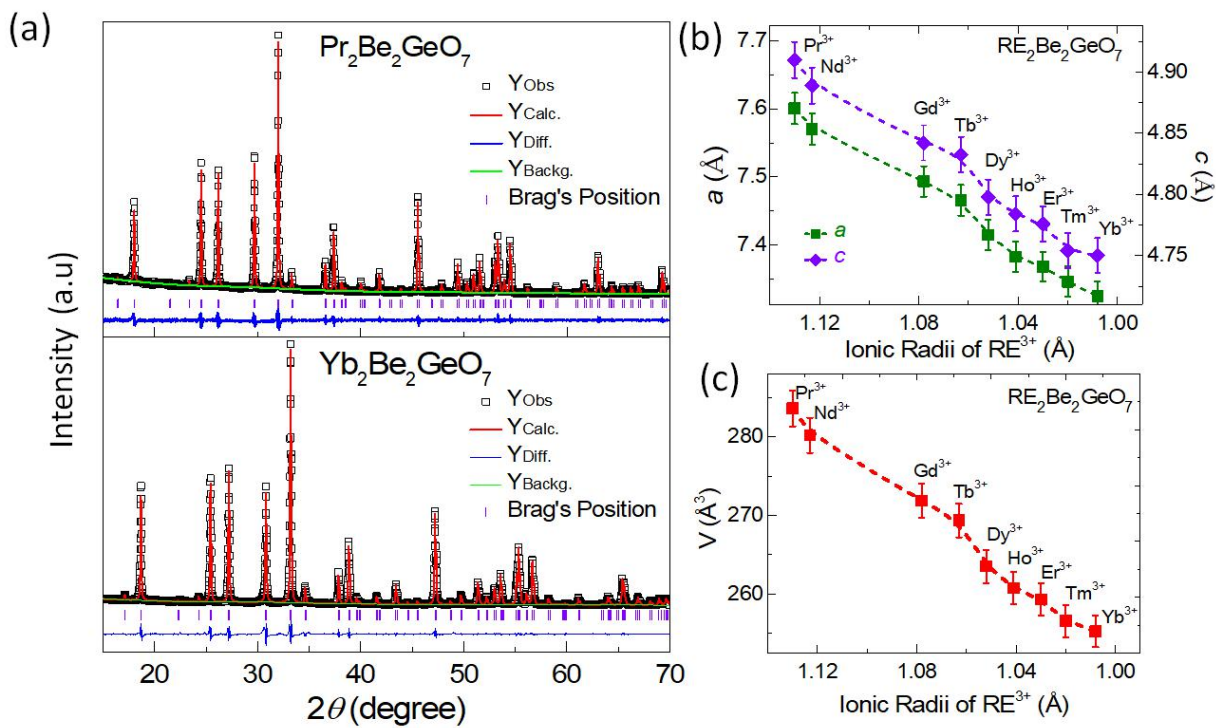

Figure 2. (a) Room-temperature XRD patterns of $RE_2Be_2GeO_7$ (Pr, Gd, Ho, Yb): black squares are experimental data; red lines show calculated patterns, blue line is differences, and violet tick represents Bragg's reflection positions. (b) The variation of lattice parameters and (c) unit-cell volume (V) as a function of ionic radii of $RE^{3+}$ ions in $RE_2Be_2GeO_7$. The dashed lines are given as guides to the eye.

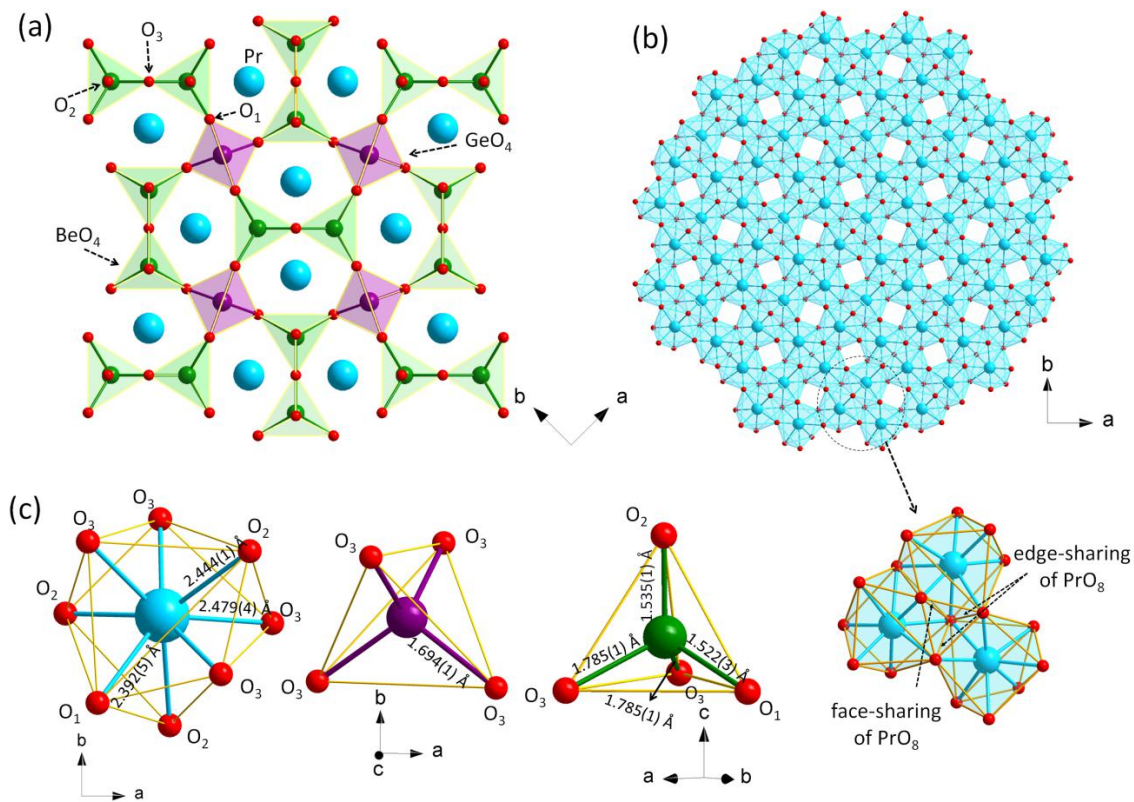

Figure 3. (a) Top view of structural framework of melilite-type $Pr_2Be_2GeO_7$, (b) The connectivity of $PrO_8$ polyhedrons in the *ab* plane. (c) The coordination environments of $PrO_8$, $BeO_4$ and $GeO_4$ polyhedron.

The schematic crystal structure of $Pr_2Be_2GeO_7$ as one representative is presented in Figure 1b, which is also applicable to the other isostructural family members. As seen, magnetic $Pr^{3+}$ ions and nonmagnetic $Be^{2+}/Ge^{4+}$ ions are located on separate layers alternatively along the *c* axis. In the nonmagnetic $[GeBe_2O_7]^{-6}$ layers, the corner- sharing $GeO_4$ tetrahedra and $Be_2O_7$ distorted tetrahedral dimers build a two-dimensional (2D) network consisted of five-membered rings [see Figure 3a]. The magnetic Pr cations with 8-fold coordination form distorted square Archimedean antiprism residing on neighbouring layers. The detailed coordination environment of oxygen ligands surrounded Pr, Be and Ge cations are shown in Figure 3b, the related interatomic distances and bond angles are provided in Table 1. For distorted $BeO_4$ tetrahedra, the Be–O distances are 1.522(3) Å, 1.535(1)Å and 1.785(1) Å, and O–Be–O bond angles are 117.0(2)°, 113.9(3)° and 104.5(1)°. The $GeO_4$ tetrahedra have four Ge–O bond lengths of ~1.694(1) Å and O–Ge–O angles ~109.2(3)°. In the $PrO_8$ polyhedron, the Pr-O bond distances are ranging from 2.392(3) Å to 2.479(4) Å, and the O–Pr–O angles are 76.1 (3)°, 108.2(3) and 146.1(1)°, respectively. Within the *ab* plane, the $PrO_8$ polyhedrons have edge and face-shared connections (see Figure 3c), magnetic $Pr^{3+}$ ions are located on topological Shastry-Sutherland lattice (SSL) pattern. The SSL layers are stacked in an eclipsed

"AAA"-type fashion along c-axis, where the inter-plane Pr-Pr distances given by $c$=4.910(1) Å are larger than the Pr-Pr diagonal bond (~3.429 Å) and square bond (~4.042 Å). Moreover, considering the well separation of magnetic layers by nonmagnetic Ge/Be-O polyhedrons, an important feature of $Pr_2Be_2GeO_7$ is that its' structure consists of magnetic $Pr^{3+}$ SSL sheets and nonmagnetic $Be^{2+}/Ge^{4+}$ layers alternatively along c axis.

Table 1. Selected bond distances, bond angles, and interplane and intraplane RE-RE distances of $RE_2Be_2GeO_7$ (RE = Pr, Nd, Gd-Yb) compounds.

| RE | Pr | Nd | Gd | Tb | Dy | Ho | Er | Tm | Yb |
|---|---|---|---|---|---|---|---|---|---|
| $REO_8$ | | | | | | | | | |
| RE–$O_1$ (Å) | 2.392(5) | 2.352(4) | 2.304(1) | 2.297(4) | 2.267(4) | 2.259(1) | 2.244(3) | 2.236(1) | 2.231(5) |
| RE–$O_2$ (Å) | 2.444(1) | 2.420(2) | 2.413(2) | 2.405(2) | 2.367(3) | 2.354(2) | 2.302(1) | 2.294(3) | 2.289(2) |
| RE–$O_3$ (Å) | 2.479(4) | 2.456(3) | 2.413(1) | 2.405(3) | 2.366(1) | 2.359(2) | 2.323(4) | 2.315(3) | 2.310(1) |
| Intraplane RE-RE (Å) | 4.042(3) | 4.027(1) | 3.988(1) | 3.972(1) | 3.942(3) | 3.927(3) | 3.919(1) | 3.907(2) | 3.896(2) |
|  | 3.429(2) | 3.409(2) | 3.372(1) | 3.360(4) | 3.339(4) | 3.327(4) | 3.322(1) | 3.312(2) | 3.303(5) |
| Interplane RE-RE (Å) | 4.910(1) | 4.889(1) | 4.842(1) | 4.833(4) | 4.798(2) | 4.784(1) | 4.776(2) | 4.754(4) | 4.750(1) |
| $O_1$–RE–$O_2$ (°) | 108.2(1) | 109.1(1) | 109.5(1) | 109.4(2) | 109.6(1) | 109.6(5) | 110.1(5) | 110.1(3) | 110.0(2) |
| $O_1$–RE–$O_3$ (°) | 146.1(1) | 146.2(1) | 146.4(4) | 146.4(2) | 1.46.6(3) | 146.7(5) | 147.5(4) | 147.5(2) | 147.5(5) |
| $O_2$–RE–$O_3$ (°) | 76.1(1) | 75.6(1) | 75.2(3) | 75.3(1) | 75.3(4) | 75.3(1) | 74.7(2) | 74.7(3) | 74.8(5) |
| $BeO_4$ | | | | | | | | | |
| Be–$O_1$ (Å) | 1.522(3) | 1.540(1) | 1.581(5) | 1.575(1) | 1.579(1) | 1.573(2) | 1.524(3) | 1.519(4) | 1.515(4) |
| Be–$O_2$ (Å) | 1.535(1) | 1.586(2) | 1.564(1) | 1.561(1) | 1.569(2) | 1.565(3) | 1.628(2) | 1.613(1) | 1.609(2) |
| Be–$O_3$ (Å) | 1.785(1) | 1.742(1) | 1.697(3) | 1.692(4) | 1.656(2) | 1.650(2) | 1.618(4) | 1.621(5) | 1.619(1) |
| $O_1$–Be–$O_2$ (°) | 117.0(1) | 115.1(1) | 114.1(5) | 114.1(1) | 114.8(1) | 114.8(1) | 117.7(2) | 117.7(5) | 117.7(2) |
| $O_1$–Be–$O_3$ (°) | 104.5(1) | 103.8(3) | 102.8(3) | 102.7(2) | 102.4(3) | 102.5(4) | 104.1(4) | 104.2(3) | 104.1(2) |
| $O_2$–Be–$O_3$ (°) | 113.9(3) | 113.9(2) | 115.9(3) | 115.9(2) | 115.9(4) | 116.0(4) | 114.2(5) | 114.1(1) | 114.2(4) |
| $GeO_3$ | | | | | | | | | |
| Ge–$O_3$ (Å) | 1.694(1) | 1.707(4) | 1.712(4) | 1.706(3) | 1.716(4) | 1.710(4) | 1.752(2) | 1.746(1) | 1.742(1) |
| $O_3$–Ge–$O_3$ (°) | 109.2(3) | 109.3(2) | 109.2(5) | 109.1(5) | 109.3(5) | 109.2(1) | 109.1(2) | 109.1(2) | 109.4(2) |

For $RE_2Be_2GeO_7$, both lattice parameters ($a$, c) and unit volume $V$ decrease almost linearly as decreased $RE^{3+}$ ionic radius, as shown in Figure 2b and 2c. These lattice variations result in the monotonic decrease of RE-RE distances. As seen in Table 1 and Figure S1, the interlayer's spacing ($d_{inter}$) between SSL layers decreases from 4.910(5) Å (RE=Pr) to 4.751(5) Å for (RE=Yb), and the nearest and next-nearest RE-RE distance ($d_1$ and $d_2$) change from $d_1$=3.430(1) Å and $d_2$=4.041(1) Å (RE=Pr) to $d_1$=3.301(3) Å and $d_2$=3.899(1) Å for (RE=Yb), respectively. On other side, the value ratio of $d_2/d_1$=1.18(1) is almost constant irrespective of the variations of RE ions. Considering that the relative strength of intradimer ($J_1$) and interdimer ($J_2$) exchanges are sensitive to the RE-RE distances ($d_1,d_2$) within the SSL plane, which play critical role in determining the magnetic behaviors. Herein, a comparison with another RE-based $RE_2BaPdO_5$ was performed, $RE_2Be_2GeO_7$ has comparable value of $d_1$ in respect to $d_1$=3.0~3.5 Å of $RE_2BaPdO_5$, but have larger $d_2/d_1$ contrast to $d_2/d_1$=1.06~1.07 of $RE_2BaPdO_5$.[32] The eclipsed stacking fashion of SSL layers in $RE_2Be_2GeO_7$ is different from $RE_2BaPdO_5$ family showing AB-type fashion. Additionally, different local coordinated environments of $RE^{3+}$ ions, resultant crystal-electric field (CEF) effect and $RE^{3+}$-O-$RE^{3+}$ superexchange pathway for above two systems can affect their magnetic behaviors. Therefore, $RE_2Be_2GeO_7$ provides an alternative platform for exploring novel quantum magnetic phases.

**Magnetic susceptibilities and isothermal Magnetizations**

Temperature ($T$) dependence of magnetic susceptibilities $\chi(T)$ for $RE_2Be_2GeO_7$ were measured under field cooling condition in an applied field $H$= 0.1 T, as shown in Figure 4a-i. Only $Tb_2Be_2GeO_7$ shows antiferromagnetic (AFM) transition characteristics with a peak at $T_N$~2.6 K (see inset of Figure 4d), none of other members exhibit magnetic transitions at $T \geq$ 2 K. The Curie-Weiss law $1/\chi(T) = (1/C)(T-\theta_{CW})$ was used to analyse the inverse susceptibilities $1/\chi(T)$ for extracting the Weiss temperature $\theta_{CW}$ and effective moment $\mu_{eff}$, the resultant magnetic parameters are listed in Table 2. The $\mu_{eff}$ is calculated by $\mu_{eff} = (3k_BC/N_A)^{1/2}$, where $k_B$ is Boltzmann constant and $N_A$ is Avogadro's constant. The Curie-Weiss fitting was performed at high temperature (100 K – 300 K) and low temperature (8 K <T< 20 K) regimes, respectively. At high temperatures, $1/\chi(T)$ follows a linear temperature dependence, the fitted value of $\mu_{eff}$ is close to theoretical free-ion value $g_J[J(J+1)]^{1/2}$. As decreased temperatures, the slope of $1/\chi(T)$ changes below ~80 K and which is especially obvious for RE=Pr, Nd and Yb compounds, reflects the variation of magnetic interactions and effective moments. This phenomenon is consistent with the magnetic contributions from different 4f-manifold excitations due to the crystal electric field (CEF) splitting, where more population of electrons will occupy the CEF ground level as decreased temperatures, leading to the reduced effective moments. For all compounds, negative values of $\theta_{CW}$ reveal the dominant AFM type interactions between $RE^{3+}$ moments. Among them, $Gd^{3+}$ is special and has half-filled 4f shell ($4f^7$, $S$=7/2, $L$=0) without orbital moments. For $Gd_2Be_2GeO_7$, the negative value of $\theta_{CW}$=-2.4 K and absence of magnetic transition ($T$ >1.8 K) suggests the dominant AFM exchange interactions between $Gd^{3+}$ moments.

The isothermal magnetization $M(H)$ curves at selected temperatures for $RE_2Be_2GeO_7$ series are shown in Figure 5a-i. Despite RE= Pr and Nd, magnetization curves at 2 K of other compounds nearly saturate at 7 T. Among them, $Gd_2Be_2GeO_7$ being a good approximation of Heisenberg antiferromagnet with $S$=7/2 moments, has saturated magnetization $M_s$ =7.05 $\mu_B$ in agreement with the expected value $g_JJ\mu_B$ = 7$\mu_B$. For RE= Dy and Yb, the saturated magnetization is around half of above value with $M_s$ ~$g_JJ\mu_B$/2, indicating of easy axis (Ising) anisotropy. For RE=Pr and Nd, the observed values of $M_s$ at 7 T are quite smaller than expected $M_S$=$g_JJ$ for free $Pr^{3+}/Nd^{3+}$ ions. This can be related to their reduced moments at low temperatures due to the low-lying CEF effects.[42] Further considering the Ising-like magnetic anisotropy, the saturated magnetization of $RE_2Be_2GeO_7$ (RE=Pr,

Nd) should be ~1.1-1.3$\mu_B$/f.u, in consistent with the experimental results. In these SSL magnetic oxides, local coordination environment of RE$^{3+}$ ions, resultant different CEF effect and competing dipolar interactions can also affect their magnetic behaviors including the magnetic anisotropic effect, low $T$ effective moment and saturated magnetization. Similar behaviors have been reported for other RE-based magnets such as triangular lattice NaRES$_2$,[22] kagome lattice RE$_3$BWO$_9$ [43] and one-dimensional Ca$_4$REO(BO$_3$)$_3$ systems.[44]

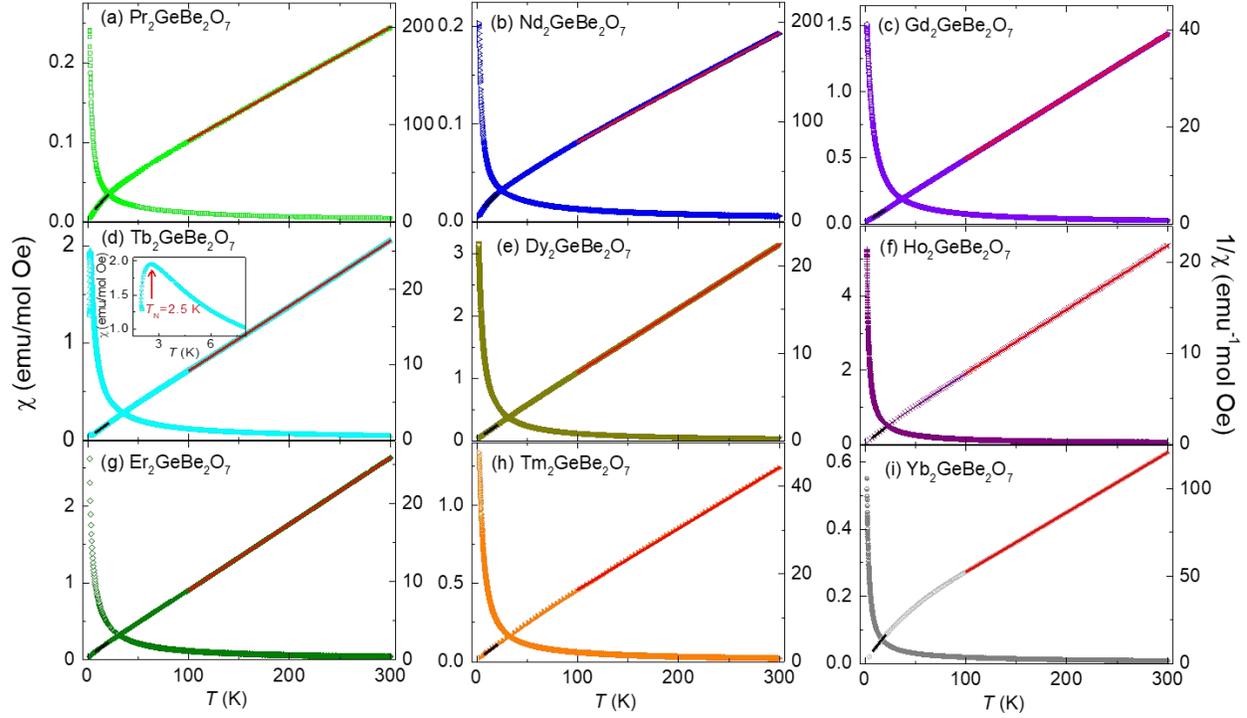

Figure 4. (a-i) Temperature-dependence of susceptibility $\chi$(T) and inverse susceptibility 1/$\chi$(T) measured at 0.1 T for RE$_2$Be$_2$GeO$_7$ (RE = Pr, Nd, Gd-Yb), respectively. The solid red and black lines show the Curie−Weiss fitting at high temperature and low temperature regimes, respectively. The inset in (d) shows the enlarged region of $\chi$(T) at low temperatures.

For Tb$_2$Be$_2$GeO$_7$, $M(H)$ curves at 2 K nearly saturate at ~2 T with $M_s$ = 4.54(5) $\mu_B$ slightly smaller than $g_JJ\mu_B$/2 with $J$=6 and $g$=3/2 for free Tb$^{3+}$ ions (4$f^8$,$^7$F$_6$) with m$_J$=±6. Moreover, below $T_N$, its magnetizations exhibit an inflection point near critical field ($H_C$) as existence of spin-flop behaviors (see figure 6a). Here, $H_C$ is defined by the peak position of derivative magnetization d$M$/d$H$, as shown in the inset of Figure 5d and Figure S3 b. The $H_C$ decreases with increased temperatures towards zero at ~2.8 K. Further combing the $T_N$ defined by the peak of magnetic susceptibilities measured at different $H$ (see Figure S3 a), the $H$-$T$ phase diagram is constructed. The contour plots using field dependent d$M$/d$H$ for the phase diagram are shown in figure 6b, the observation of spin-flop transition from AFM to canted-AFM state below $T_N$ reveals the evolution of magnetic configurations. The field induced successive multiple metamagnetic transitions are reported in SSL TbB$_4$ system,[30] as comparisons, further study on field-induced metamagnetic transition based on Tb$_2$Be$_2$GeO$_7$ single crystals are desirable.

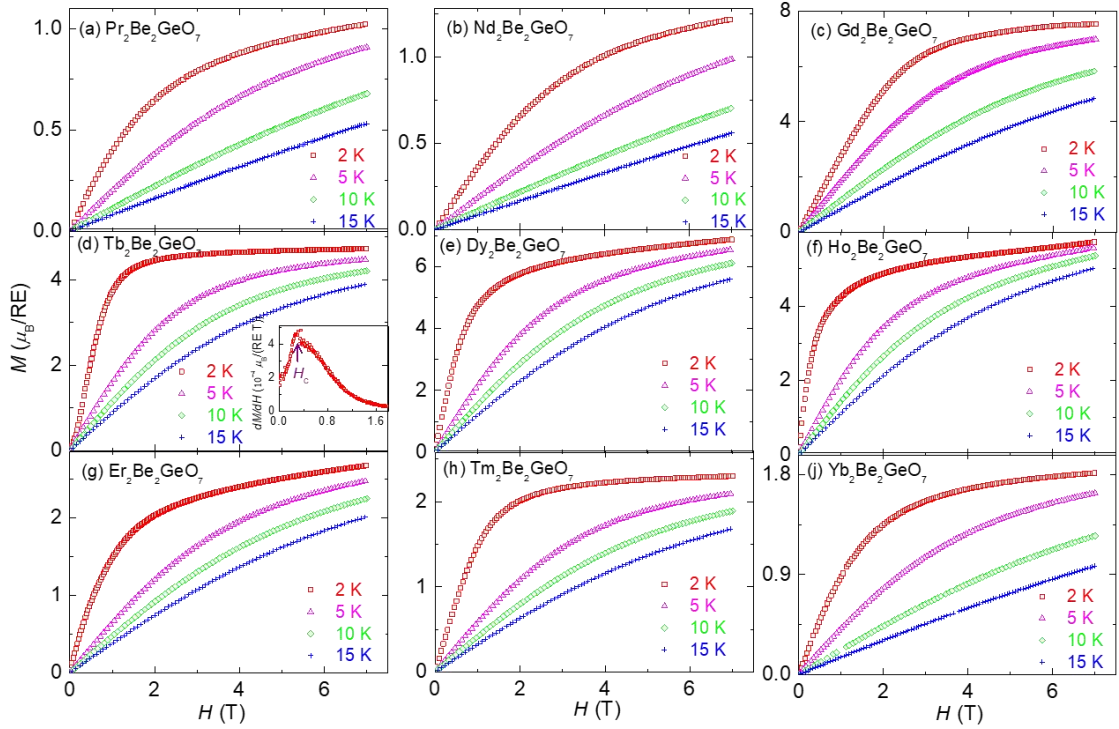

Figure 5. Field-dependence of magnetization $M(H)$ curves at selected temperatures for RE$_2$Be$_2$GeO$_7$ (RE = Pr, Nd, Gd-Yb). The inset in (d) shows the derivative magnetization d$M$/d$H$ at 2 K.

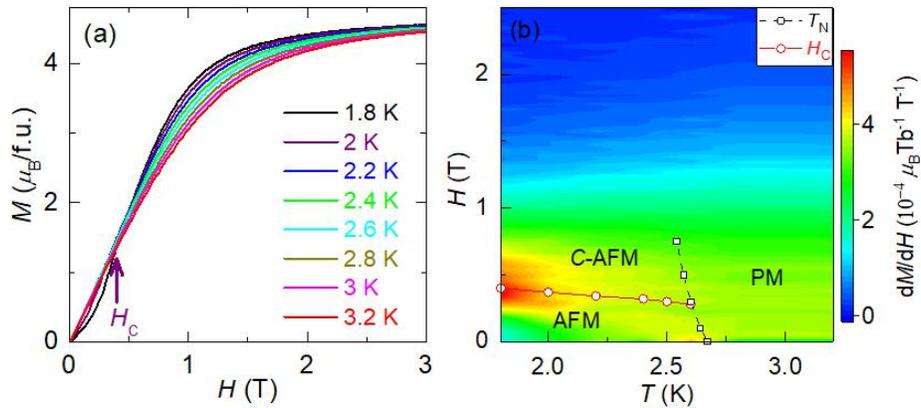

Figure 6. (a) The isothermal $M(H)$ curves for Tb$_2$Be$_2$GeO$_7$ measured at $T \leq 3.2$ K, (b) The constructed field-temperature ($H$-$T$) phase diagrams based on d$M$/d$H$ and $M(T)$ data.

Table 2. Magnetic parameters obtained from the fitting of χ(T) by the Curie–Weiss law, Curie–Weiss temperatures ($\theta_{CW}$) and effective magnetic moments ($\mu_{eff}$), and the effective moment ($\mu_{fi}$) expected for free ions calculated by $g[J(J+1)]^{1/2}$ for for $RE_2Be_2GeO_7$ (RE = Pr, Nd, Gd-Yb).

| RE | High T fit (K) | $\theta_{CW}$ (K) | $\mu_{eff}$ ($\mu_B$) | Low T fit (K) | $\theta_{CW}$ (K) | $\mu_{eff}$ ($\mu_B$) | $\mu_{fi}$ ($\mu_B$) |
|---|---|---|---|---|---|---|---|
| Pr | 100-300 | -43.3 (1) | 3.70(1) | 8-20 | -3.61(3) | 2.60(1) | 3.58 |
| Nd | 100-300 | -45.6(1) | 3.82(1) | 8-20 | -6.25(5) | 2.76(2) | 3.62 |
| Gd | 100-300 | -4.09(5) | 7.88(2) | 8-20 | -2.41(2) | 7.79 (1) | 7.94 |
| Tb | 100-300 | -5.95(2) | 9.63(1) | 8-20 | -2.16(2) | 9.06(3) | 9.72 |
| Dy | 100-300 | -6.87(6) | 10.45(2) | 8-20 | -1.74(2) | 9.92(1) | 10.6 |
| Ho | 100-300 | -10.58(1) | 10.66(2) | 8-20 | -1.05(2) | 9.63(1) | 10.60 |
| Er | 100-300 | -4.11 | 9.70(1) | 8-20 | -3.45(1) | 9.31(2) | 9.6 |
| Tm | 100-300 | -17.5 | 7.57(1) | 8-20 | -2.12(1) | 6.78(1) | 7.6 |
| Yb | 100-300 | -54.8(1) | 4.65(2) | 8-20 | -0.91 | 3.18(1) | 4.54 |

Table 3. Dipolar interaction (D) and nearest-neighbour exchange interaction energy ($J_{nn}$) for $RE_2Be_2GeO_7$ (RE = Pr, Nd, Gd- Yb), where $J_{nn}$ is estimated in the mean-field isotropic approximation.[45]

| RE | $D_{intralayer}$ (K) | | $D_{intralayer}$ (K) | $J_{nn}$ (K) using J | $J_{nn}$ (K) using $J_{eff}$ =1/2 |
|---|---|---|---|---|---|
| | $D_{(square\ bond)}$ | $D_{(square\ bond)}$ | | | |
| Pr | 0.0639(1) | 0.105(1) | 0.0358(2) | -0.108(1) | -2.88(2) |
| Nd | 0.0736(2) | 0.121(1) | 0.0406(1) | -0.152(1) | -5 |
| Gd | 0.600(1) | 0.981(1) | 0.332 (2) | -0.0918(1) | N/A |
| Tb | 0.817(3) | 1.35(1) | 0.456(3) | -0.0309(1) | -1.73(1) |
| Dy | 1.002(2) | 1.66(2) | 0.554(2) | -0.0164(1) | -1.39(2) |
| Ho | 0.958(3) | 1.56(1) | 0.528(1) | -0.0088(1) | -0.84(2) |
| Er | 0.897(2) | 1.49(1) | 0.497(2) | -0.0325(1) | -2.76(1) |
| Tm | 0.483(2) | 0.790(1) | 0.267(1) | -0.0303(1) | -1.69(1) |
| Yb | 0.106(2) | 0.175(2) | 0.059(1) | -0.035(1) | -0.73(1) |

The $RE^{3+}$ ions can be divided into two categories, Kramers ions containing odd 4f electrons ($Nd^{3+}$, $Dy^{3+}$, $Er^{3+}$ and $Yb^{3+}$) and non-Kramers ions with even numbers of 4f electrons ($Pr^{3+}$, $Tb^{3+}$, $Ho^{3+}$ and $Tm^{3+}$). Among them, $Gd^{3+}$ has half-filled 4f shell with Heisenberg-like anisotropy, the effective moment is $S_{eff}$=7/2. For other RE-based compounds, effective local spin moment for Kramer's doublet can be described by $S_{eff}$ =1/2 protected by time-reversal symmetry. For non-Kramers ions, the $S_{eff}$ = 1/2 ground state can be formed for low crystal-field symmetries, where crystal symmetry cannot provide enough symmetry operations protecting the degeneracy of the crystal-field levels. Here, due to the absence of inelastic neutron spectroscopy (INS) data of $RE_2Be_2GeO_7$, the CEF and spin anisotropy cannot be accurately determined. Here, the strength of superexchange interactions between local $RE^{3+}$ moments are estimated by the mean-field approximation using $J_{nn} = 3k_B\theta_{CW}/zS_{eff}(S_{eff}+1)$,[45] where $S_{eff}$ denominates the total spin number, and z is the number of nearest-neighbor spins (here z=5). Considering the strong spin-orbit coupling (SOC) and different spin types of $RE^{3+}$, both quantum number $J$ and $S_{eff}$ are used to estimate the nearest-neighbor exchange $J_{nn}$ (see Table 3). The dipolar interaction $D$ may be calculated by $D = \mu_0\mu_{eff}^2/4\pi(r_{nn})^3$, where $r_{nn}$ is the distance between adjacent $RE^{3+}$ in the SSL layers. Using the low $T$ fitted $\mu_{eff}$, the sizes of dipolar interactions are also listed in Table 3. As seen, the intralayers' dipolar energy is 1.8~2.9 times larger than the interlayers' ones. As an example, for $Yb_2Be_2GeO_7$, the interlayer dipole interaction ($D_{interlayer}$~0.059 K) is ~ 2.4 times weaker than the average value of intralayers' interaction ($D_{intralayer}$~0.14 K). The simultaneous mediation by exchange/dipolar interactions, orbital hybridization and presence of varies spin types (Ising, Heisenberg, and planar XY) in $RE_2Be_2GeO_7$ melilites would be important factors, which should be considered to determine the magnetic interactions and magnetic ground states. Moreover, considering similar geometric networks of RE-O-RE or RE-O-O-RE superexchange pathway in this family, the highest magnetic ordering temperature of $Tb_2Be_2GeO_7$ reveals other factors like orbital hybridization inherent to different 4f orbit on magnetic exchange interactions should be considered. Similar behavior in Quasi-One-Dimensional RE-based magnets $Ca_4LnO(BO_3)_3$ has also been reported,[44] where only $Ca_4TbO(BO_3)_3$ compound undergo magnetic transition above 2 K ($T_N$ = 3.6 K).

**Magnetocaloric effect**

The RE-based compounds are quite suitable as magnetic refrigeration materials, which can show large magnetocaloric (MC) effect due to the large number of unpaired spins and great magnetic moments. In Figure 5, four members for RE = Gd, Tb, Dy, Ho are identified to have relatively large magnetizations, then we focus on the MC studies on above compounds. Magnetic entropy changes ($\Delta S_m$) were calculated from isothermal process of magnetization by employing Maxwell's relation: [46] $\Delta S_m = \int_{H_i}^{H_f} \left(\partial M/\partial T\right)_H dH$, where $H_i$ (usually $H_i$=0) and $H_f$ represent the initial and final values of magnetic field, respectively. The obtained field dependence of $-\Delta S_m$ at 2 K are shown in Figure 7. In this family, $Gd_2Be_2GeO_7$ shows the optimal MCE in large field regions (5 T ≤ $H$ ≤ 7 T) with maximum $-\Delta S_m$ =54.8 J K$^{-1}$ kg$^{-1}$ at 2 K, while $Dy_2Be_2GeO_7$ has the best MCE at low fields $H$ ≤ 2 T, where 2 T is the largest field attainable with permanent magnets. This MCE at low field is more attractive for practical applications free of cryogens.

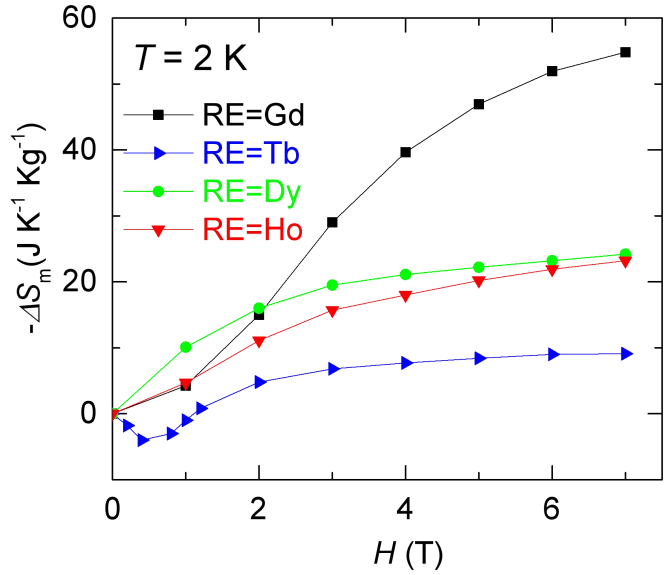

Figure 7. The field dependence of magnetic entropy change $\Delta S_m$ for $RE_2Be_2GeO_7$ (RE = Gd,Tb,Dy,Ho) at 2 K.

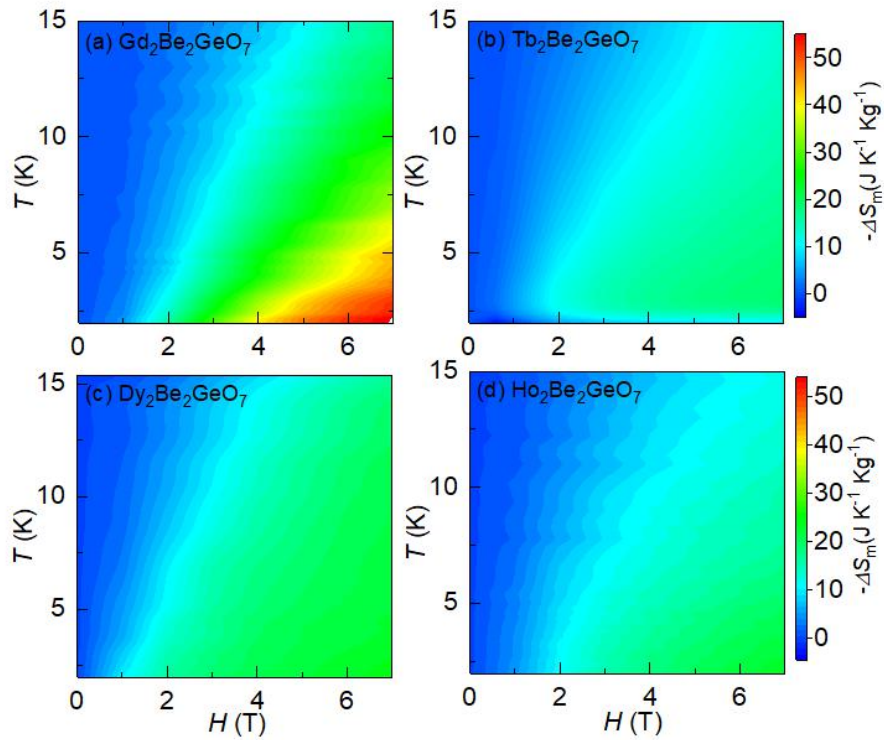

Figure 8. (a-d) The contour plots of obtained $-\Delta S_m$ as function of temperature (2 K - 15 K) and field for $RE_2Be_2GeO_7$ (RE = Gd,Tb,Dy,Ho).

Moreover, the contour plots of obtained $-\Delta S_m$ as function of temperature (2 K- 15 K) and field for $RE_2Be_2GeO_7$ (RE = Gd, Tb, Dy, Ho) melilites are given in Figure. 8. As compared to the benchmark MC material $Gd_3Ga_5O_{12}$ (GGG) or $Dy_3Ga_5O_{12}$ (DGG), the obtained MCE of $RE_2Be_2GeO_7$ (RE=Gd, Dy) is impressive, high field ($\Delta H$= 7 T) $\Delta S_m$ =54.8 J $K^{-1}$ $Kg^{-1}$ of $Gd_2Be_2GeO_7$ is much larger than that of GGG ($-\Delta S_m$ = 38.5 J $K^{-1}$ $Kg^{-1}$) at 2 K.[35,36] In lower field change $\Delta H$=2 T, MCE performance of $Dy_2Be_2GeO_7$ (16.1 J $K^{-1}$ $kg^{-1}$) also surpasses DGG (11.0 J $K^{-1}$ $kg^{-1}$) and GGG (14.6 J $K^{-1}$ $kg^{-1}$) at 2 K, see Table S1. Additionally, $Tb_2Be_2GeO_7$ exhibits maximum $-\Delta S_m$ at 2.8 K for all fields, around the ordering temperature $T_N$~2.5 K, then $-\Delta S_m$ fall off on both side temperatures, as observed in $RE_3CrGa_4O_{12}$ (RE = Tb, Dy, Ho).[36,47] The MCE of $Dy_2Be_2GeO_7$ has remarkable values and weak temperature dependence between 2 and 15 K. The large MCE make $RE_2Be_2GeO_7$ as competitive magnetic refrigeration materials at liquid Helium temperatures, further optimised magnetocaloric efficiency at different temperature and field regimes could be realized by partial chemical substitutions or solid-state solutions such as $Gd_{2-x}Dy_xBe_2GeO_7$.

According to the SSL models, its magnetic ground states strongly depend on the intra- and inter-dimer couplings ratio $J_2/J_1$ strength, which is experimentally confirmed by the well-studied spin-1/2 dimerized SSL magnets $SrCu_2(BO_3)_2$.[14-17] As comparison, magnetic ground states of RE-based SSL magnets should also be affected by the anisotropic spin exchange interactions due to the strong spin-orbital-entanglement and CEF effects. As an example, the present $Yb_2Be_2GeO_7$ will provide comparative study on magnetic ground state between $Cu^{2+}$ ($S$=1/2) and $Yb^{3+}$ ($J_{eff}$=1/2) SSL based systems. The varies of different spin types (Ising, Heisenberg, and planar XY) in $RE_2Be_2GeO_7$ melilites would be attractive for discovering diverse novel magnetic ground states, such as spin liquid state, dimers state, etc. New theoretical model combing the above effects and extensive ultra-low-temperature experimental studies based on high quality single crystals will be significant to elucidating the ground state of these systems, and the availability of $RE_2Be_2GeO_7$ single crystals will be highly desirable to unveil anisotropic magnetocaloric effect.

## ■ CONCLUSIONS

A new family of RE-based magnets $RE_2Be_2GeO_7$ (RE = Pr, Nd, Gd-Yb) were synthesized, where $RE^{3+}$ ions are located on the SSL within the ab plane. Structurally, the SSL layers exhibit the eclipsed AAA-type stacking in comparison with the AB-type fashion of $RE_2BaZnO_5$. In this family, most $RE_2Be_2GeO_7$ (RE = Pr, Nd, Gd-Yb) compounds except of RE = Tb show no magnetic order down to 2 K, where $Tb_2Be_2GeO_7$ exhibits AFM transition with Néel temperature $T_N$~ 2.5 K and field-induced spin flop behaviors ($T<T_N$). The calculated magnetic entropy change $\Delta S_m$ reveal large magnetocaloric effect (MCE) for $RE_2Be_2GeO_7$ (RE =Gd, Dy) in liquid helium temperature regimes, $Gd_2Be_2GeO_7$ shows the optimal MCE (54.8 J $K^{-1}$ $Kg^{-1}$) at 7 T and $Dy_2Be_2GeO_7$ has the maximum

MCE (16.1 J K$^{-1}$ kg$^{-1}$) at 2 T as suitable materials for magnetic refrigeration. Moreover, the present SSL magnets incorporating strong spin orbit coupled 4f ions with $J_{eff}$=1/2 spin enable the possibility to explore novel quantum magnetic phases beyond the SrCu$_2$(BO$_3$)$_2$ as typical $S$=1/2 Cu$^{2+}$-based SSL magnet.

## ■ ASSOCIATED CONTENTS

**Supporting information.**

The room-temperature XRD spectra and refined lattice parameters of RE$_2$Be$_2$GeO$_7$, magnetic results of Tb$_2$Be$_2$GeO$_7$ are provided.

## ■ AUTHOR INFORMATION

**Corresponding Author**

**Zhaoming Tian –**_Wuhan National High Magnetic Field Center and School of Physics, Huazhong University of Science and Technology, Wuhan, 430074, P. R. China;_ orcid.org/0000-0001-6538-3311; Email:tianzhaoming@hust.edu.cn

**Authors**

**Malik Ashtar –**Wuhan National High Magnetic Field Center and School of Physics, Huazhong University of Science and Technology, Wuhan, 430074, P. R. China

**Yuming Bai –**Wuhan National High Magnetic Field Center and School of Physics, Huazhong University of Science and Technology, Wuhan, 430074, P. R. China

**Longmeng Xu –**Wuhan National High Magnetic Field Center and School of Physics, Huazhong University of Science and Technology, Wuhan, 430074, P. R. China

**Zongtang Wan –**Wuhan National High Magnetic Field Center and School of Physics, Huazhong University of Science and Technology, Wuhan, 430074, P. R. China

**Zijun Wei –**Wuhan National High Magnetic Field Center and School of Physics, Huazhong University of Science and Technology, Wuhan, 430074, P. R. China

**Yong Liu –**School of Physics, Wuhan University, Wuhan 430072, Peoples R China.

**Mohsin Ali Marwat –**College of Materials Science and Engineering, Huazhong University of Science and Technology, Wuhan 430074, PR China

**Notes**

The authors declare no competing financial interest.

## ◼ ACKNOWLEDGEMENTS

We acknowledge financial support from the National Natural Science Foundation of China (Grant No. 11874158 and 11604281), and Fundamental Research Funds for the Central Universities (Grant No. 2018KFYYXJJ038 and 2019KFYXKJC008). We would like to thank the staff of the analysis centre of Huazhong University of Science and Technology for their assistance in structural characterization and analysis.

## ◼ REFERENCES